\documentclass{mem}
\usepackage{txfonts}
\usepackage{balance}
\usepackage{graphicx}
\usepackage[a4paper]{hyperref}
\idline{75}{282}
\begin{document}
\def\teff{$T\rm_{eff }$}
\def\kms{$\mathrm {km s}^{-1}$}
   
\title{The nature of gas and stars in the circumnuclear regions of AGN: a chemical approach}

   \subtitle{}

\author{
P. \,Rafanelli\inst{1},
L. \, Vaona\inst{1},
R. \, D'Abrusco\inst{1},
S. \, Ciroi\inst{1},
V. \, Cracco\inst{1}}
 
\institute{Universit\'a di Padova, Dipartimento di Astronomia, 
vic. Osservatorio 3, 35122 Padova, Italia.
\email{dabrusco@unipd.it}}
\authorrunning{Rafanelli}

\titlerunning{The nature of gas and stars in the circumnuclear regions of AGN}

\abstract{Aim of this communication is to describe the first results of a work-in-progress regarding 
the chemical properties of gas and stars in the circumnuclear regions of nearby galaxies. 
Different techniques have been employed to estimate the abundances of chemical elements 
in the gaseous and stellar components of nuclear surroundings in different classes of galaxies 
according to the level of activity of the nucleus (normal or passive, star forming galaxies and AGNs).}

\maketitle

\section{Introduction}

A recent comparative study of the features of the stellar component of the circumnuclear ($r \sim 5pc$)
regions, carried out on three different samples of galaxies, namely 1302 star-forming galaxies (SFG), 1996 active galaxies and
2000 normal galaxies (passive galaxies) by \cite{rafanelli2009}, has shown that SFGs are characterized by a spectral continuum 
distribution dominated by O, B stars. On the contrary, O and B stars are absent in active and normal galaxies
which show a similar continuum in the optical spectral range, but a slightly different depth of the 4000$\AA$ break, deeper in the spectra of normal galaxies (NG) than in the spectra of active galaxies. This difference in deepness of the 4000$\AA$ break is  a clear indication that there are more stars of spectral type A, namely tracers of a recent star formation history, in the circumnuclear regions of active galactic nuclei (AGN) than in the same regions of not-active galaxies. This finding reinforces the scenario
according to which galactic activity is correlated to stellar formation in the proximity of the central engine of AGNs.
Aim of this work is the characterization of the chemical composition of the gaseous
component of the circumnuclear regions in both SFG and AGN. We have focused our attention  on two different
questions:
\begin{itemize}
\item Is gas in which star formation is occurring in SFG rich in heavy elements (taking the Sun chemical composition
as reference)?
\item  Is the chemical composition of gas ionized by the central engine of AGN different from that of SFG?
\end{itemize}   
In our work, we have applied the CLOUDY photoionization code \cite{fer98} to the AGN and 
SFG samples on a grid of models spanning a large area of the parameter space and obtained the corresponding
observed emission lines. In particular, we have concentrated our attention to Type 2 AGN, namely the Seyfert 2 
galaxies in the redshift range  $[0.04, 0.08]$, in order to avoid the broad emission lines arising in the dense gaseous clouds
close to the nucleus of the AGN and hidden by the torus in Seyfert 2 galaxies (S2G). Our analysis allows us to derive
the features of the gas components in S2 and SF galaxies with density and opacity comparable to those of the 
gaseous components in observed galaxies.

\section{The samples of galaxies}

The three different samples of galaxy spectra studied in this work have been extracted from the SDSS DR7 
spectroscopic dataset and classified on the basis of a set of classical spectroscopic diagnostic diagrams which 
exploit the flux ratios of spectral emission lines to determine the presence of AGN activity in the spectrum of the nuclear 
region of the galaxies. These samples have all been selected from a larger dataset composed of all galaxies observed 
spectroscopically and with redshift comprised in the range [0.04, 0.08], with signal-to-noise $S/N >  5$ in the spectral continuum at 
$\lambda = 5500 \AA$ with the additional requirement of a signal-to-noise ratio $S/N > 3$ in the strong emission lines 
$\mbox{[OIII]}\lambda 5007$, $\mbox{[OI]}\lambda 6300$, $\mbox{[OII]}\lambda 3727$, $\mbox{H}\beta$, $\mbox{H}\alpha$, $\mbox{[NII]}\lambda 6584$, 
$\mbox{[SII]}\lambda\lambda 6717\ 6731$, when observed. The selection in redshift has been performed in order to minimize the effect on the spectra of 
the fixed aperture of spectroscopic fibers used for SDSS observations (see \cite{kewley2006} for comparison) and allow the smallest possible contamination 
by stellar light emitted in the regions of the galaxies far outside from the nucleus. The first step of the selection process has 
been the separation of the sources without emission lines from the emission line galaxies. The 2000 galaxies with highest S/N ratios 
of this class have been chosen as members of the passive galaxies sample. The remaining emission line galaxies have been 
split in two subsamples, the first corresponding to the sources showing the signs of the presence of an AGN and the second containing SFG, 
Liners and composite galaxies by applying the diagnostic diagram based on the Oxygen lines flux ratios \cite{vaona2009}, using the fluxes 
measured by the SDSS and the empirical relation in the$\mbox{[OI]}\lambda 6300/\mbox{[OIII]}\lambda 5007$ vs $\mbox{[OII]}\lambda 3727/\mbox{[OIII]}\lambda 5007$ 
diagram derived by \cite{vaona2009}. The spectra of the galaxies selected as AGN-hosting sources in the previous step have been retrieved 
by the spectroscopic SDSS database and corrected for galactic reddening using the values of the extinction provided by the Nasa Extragalactic Database 
(NED) web-service\footnote{At the website {\it http://irsa.ipac.caltech.edu/applications/DUST/}}, based on the maps and tools discussed in \cite{schlegel1998}. 
Then, they have been reduced to the rest-frame using the value of the redshift as measured by the SDSS spectroscopic pipeline. The stellar continuum of 
the spectra of these presumptive Seyfert galaxies has been evaluated using STARLIGHT (\cite{bruzual2003}), and subtracted 
from the observed spectra in order to retain the emission and absorption features of the spectra. Afterthen, the spectral parameters of all emission 
lines with $S/N > 5$ have been re-measured using an original method performing a multi-gaussian fit of the the emission lines 
developed by L. Vaona during his PhD thesis (\cite{vaona2009}). The galaxies showing in their spectra  $\mbox{H}\alpha$ and 
$\mbox{H}\beta$ line profiles broader than the $\mbox{[OIII]}\lambda 5007$ profile have been rejected, and the remaining sample
has been classified using the classical spectroscopic diagnostic diagrams known as Veilleux-Osterbrock diagrams \cite{veilleux}. 
The last step of the selection process of the S2G galaxies has been the extraction of the galaxies which satisfy the empirical relations 
in the  $\mbox{[NII]}\lambda 6584/\mbox{H}\alpha$ vs $\mbox{[OIII]}\lambda 5007/\mbox{H}\beta$, $\mbox{[SII]}\lambda 6717/\mbox{H}\alpha$ 
vs $\mbox{[OIII]}\lambda 5007/\mbox{H}\beta$ and $\mbox{[OI]}\lambda 6300/\mbox{H}\alpha$ vs $\mbox{[OIII]}\lambda 5007/\mbox{H}\beta$ 
diagrams as determined in (\cite{kewley2006}) for a sample of galaxies observed spectroscopically in the SDSS DR4:

\begin{eqnarray}
\frac{0.61}{[\log \frac{\mbox{[NII]}}{\mbox{H}\alpha} - 0.47]} + 1.19 < \log \frac{\mbox{[OIII]}}{\mbox{H}\beta}\\
\frac{0.72}{[\log \frac{\mbox{[SII]}}{\mbox{H}\alpha} - 0.32]} + 1.30 < \log \frac{\mbox{[OIII]}}{\mbox{H}\beta}\\
\frac{0.73}{[\log \frac{\mbox{[OI]}}{\mbox{H}\alpha} - 0.59]} + 1.33 < \log \frac{\mbox{[OIII]}}{\mbox{H}\beta}
\end{eqnarray}

The SFG galaxies have been extracted from the original sample using the Oxygen diagram and then applying again the appropriate relations
in the Veilleux-Osterbrock diagnostic diagrams to the line fluxes as measured by the SDSS pipeline. To summarize, the final numbers of members 
of the NG, SFG and S2G galaxy samples used for the following analysis are 2000, 1302 and 1996 respectively. 

\section{Spectral continua}

\begin{figure}
\centering
\includegraphics[width=6cm]{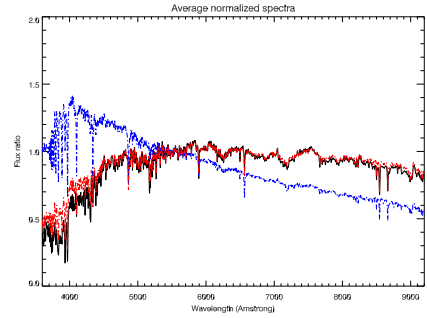}
\caption{Normalized average spectra for NGs (black curve), S2G galaxies (red curve) and SFGs galaxies (blue curve).}
\label{avgspectranorm}
\end{figure}

\begin{figure}
\centering
\includegraphics[width=6cm]{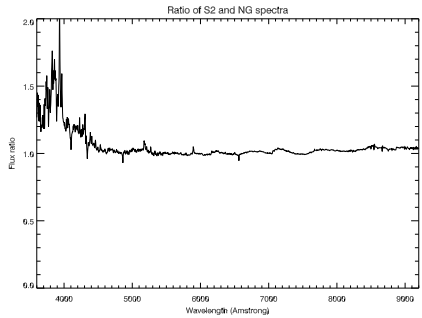}
\caption{Ratio of the normalized average spectra of NGs and S2G galaxies.}
\label{ratiospectra}
\end{figure}

The first step of the analysis has been the determination of the properties of the stellar population in the close surroundings of the galactic nucleus.
The spectra of the galaxies belonging to each of the three samples described in the previous section have been corrected for galactic reddening 
using the extinctions provided by the NED service and, similarly at what has been done for the selection of the S2G galaxies, reduced to the rest 
frame using the spectroscopic redshift measured by the SDSS. Then, all spectra have been realigned to match the dispersion of 1 \AA/pixel in 
the spectral range $[3600\AA, 9200 \AA]$. At this point, STARLIGHT has been used to produce the template spectra accounting for 
the purely stellar emission in the nuclear regions of the galaxies. This code performs spectral synthesis on observed spectra, consisting in the 
decomposition of the spectrum in terms of a convenient superposition of a base of simple stellar populations of various ages and metallicities. 
The collection of stellar populations employed to perform the fits of the input spectra in this work consists in a set of 92 stellar populations 
extracted from the Bruzual-Charlot library of evolutionary synthesis models described \cite{bruzual2003} with 23 equally spaced distinct ages 
in the interval $[10^{6}, 1.3\cdot10^{10}]$ years and 4 distinct metallicity values (Z = $4\cdot10^{-3}, 8\cdot10^{-3}, 2\cdot10^{-2}, 5\cdot10^{-2}$).
The template spectra, obtained by STARLIGHT, of all galaxies of the three samples have been normalized to the spectral flux measured at $\lambda = 5500$\AA, 
since this region of the spectrum is devoid of emission and absorption lines, thus being a good approximation of the continuum component 
of the spectrum. In figure \ref{avgspectranorm}, the comparison of the normalized average spectra for the three different families is shown 
(blue is used for SFG, red for S2G galaxies and black for NGs). While the average spectrum of SFGs galaxies is significantly higher than others 
in the shorter wavelengths region, the behaviour of the spectra of S2G and NGs galaxies is very similar. The ratio of the average normalized 
spectra of S2G galaxies and NGs is shown in figure \ref{ratiospectra} in order to highlight the differences between their shapes. The average 
spectrum of S2G galaxies appears to be systematically higher than the average spectrum of the NGs at wavelengths smaller than $\lambda = 5500 \AA$, 
thus indicating the presence of a small fraction of young stellar populations in the nuclear regions of galaxies that harbor a Seyfert 2 nucleus which are not 
found in the surrounding of the nuclei of no-emission lines galaxies.  

\section{Metallicity of the three samples of galaxies}

Three different methods have been used to estimate the metallicity of the interstellar matter and stellar populations in the central 
regions of galaxies of the three samples here considered. For Seyfert 2 galaxies, a photoionization model for the gas 
in the narrow line region of the AGN (\cite{vaona2009}) has been employed, while the content of metals of central regions of SFs 
and normal galaxies has been evaluated using two empirical techniques, the P method (\cite{pilyugin2005}) and 
the $Mg_{2}$ index method (\cite{bender1993}) respectively. The last method, namely the empirical correlation between the $Mg_{2}$
index and the metallicity, has been employed also to estimate the metallicity of S2 and SFs galaxies and compared with the metallicity
provided by the other methods. In both cases, the circumnuclear stellar populations have been reconstructed using STARLIGHT. 
In the following sections these methods and the results of their application to the galaxy samples considered will be described.

\subsection{Photoionization models}

There are so many physical processes involved in the ionized gas emitting region of AGNs that building a model able to reproduce the observed spectrum is a great challenge. Following (\cite
{netzer08}), there are five main aspects to be taken into account for a correct modeling of the ionized gas physical processes: photoionization and radiative 
recombination, thermal balance, ionizing spectrum, gas chemical composition and clouds and/or filaments distribution. In the thermal balance we find the 
mechanical heating and the role of dust which is always present in the NLRs. We have excluded from our analysis the gas kinematics and winds and gas confinement
mechanisms. A code such as CLOUDY \cite{fer98} is able to solve numerically the ionization and thermal structure of a single cloud so it can be used to explore different parameters values, 
to build more sophisticated models such as  composite models with more clouds and/or different geometries and time-dependent models. CLOUDY works with two defaults 
geometries, open and close respectively. The first one is based on simple photoionization field through a plane parallel slab with a defined ionized column density, while in the 
second case the gas is distributed all around the source. The models were made using CLOUDY code version $06.02$ \cite{fer98}. We have assumed open geometry for all models.
Once the ionizing spectrum is defined (no-thermal power-law in our case) the input parameters are the density $N_{e}/cm^{-3}$, the ionized column density $N_{c}(H^{+})/cm^{-2}$, 
the ionization parameter $U$  and the metal abundances,$Z/Z_\odot$. The ionization parameter expresses the level of ionization of photoionized gas and is proportional to the ratio of the 
ionizing photon density over the gas density according to the relation: 

\begin{equation}
 U= \frac {Q(H)} {4 \pi r^{2} c N_{H}}
\end{equation}
\label{eq:U}

where c, the speed of light, is introduced to make U dimensionless. We built two kinds of photoionization models: single cloud models and composite models, the latter being
based on a combination of two different single-cloud models. In the literature two fundamental approaches regarding composite models can be found: a dense cloud inside a 
medium with low density (\cite{bws96}, or combining separately two clouds (\cite{ks97}). The evaluation of the model parameters on a grid of models is the only realistic way to compare 
a great number of observed spectra to the models. Since our aim is to define the mean physical parameters which characterize the NLRs using spectra in the visible range, a particular 
attention is put on the metallicity because this parameter is important in the investigation of the nature of gas and its origin. We have generated a great number of models in 
order to fit all the observed lines. The well known and useful method of the line ratio diagrams employs few lines and so has limited separation power, so that if we reproduce 
all the measured lines the line ratio diagrams are automatically satisfied by a large number of models. In these cases, parameters values are obtained by averaging the 
different models. Another important result obtained by this method is the extension toward the ultraviolet (UV) and near infrared (NIR) wavelengths of the electromagnetic spectrum. 
A $\chi^2$ test has been employed to compare the models with the observed fluxes.

\subsubsection{One cloud models and the input parameters}

\begin{figure}
 \centering
 \includegraphics[width=6cm]{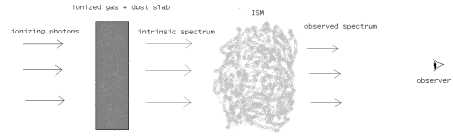}
 \caption{Basic scheme of the photoionization models.}
 \label{fig:base_model}
\end{figure}

In order to explore the 5-dimensional parameter space we built, a set of single cloud models with the values for the parameters reported in Table \ref{tab:input_par}. The ionizing spectra were 
assumed to be power-laws with $\nu^{\alpha}$ in the range $10$ $\mu$$m$ - $50$$Kev$, a cut-off at low energies ($\nu^{5/2}$) and a cut-off at 
high energies ($\nu^{-2}$). This choice is in agreement with the results showed in (\cite{ks97}) and (\cite{gds04}), and supported by X band observations 
(see \cite{mai07}). In order to assure consistency, the column density values were checked and the simulation was interrupted when the number of transmitted 
ionizing photons was below $1$\%; otherwise, the simulation stops when the final temperature reaches $4000K$. The basic model is 
summarized in Figure \ref{fig:base_model}: the ionizing photons cross a slab with a fixed density, column density and dust to gas ratio (D/G) (the distance of the source is not 
required as a free parameter since the ionization parameter is fixed). Then the intrinsic spectrum is calculated using CLOUDY: the calculation stops when either the 
fixed required column density has been achieved or the temperature in the simulation has fallen below $4000$K. We assume that the intrinsic spectrum is reddened 
by the ISM, so when the models are compared to the observed spectra it is necessary to correct for extinction.  

\begin{table}[width=6cm]
  \caption{Input parameter values. Note that $U$, $N_{e}$ and $N_{c}(H+)$ values are reported in logarithmic notation.}
  \begin{tabular}{@{}lcccccc}
\hline
$Z/Z\sun$ & $\alpha$ & $U$ & $N_{e}$ $cm^{-3}$ & $N_{c}(H+)$ $cm^{-3}$ \\ 
\hline
0.5 & -1.9 & -3.6 & 1 & 19 \\ 
1 & -1.6 & -3.2 & 1.5 & 19.4 \\ 
1.5 & -1.3 & -2.8 & 2 & 19.8 \\ 
2 & -1 & -2.4 & 2.5 & 20.2 \\ 
2.5 &  & -2 & 3 & 20.6 \\ 
3 &  & -1.6 & 3.5 & 21 \\ 
3.5 &  &  & 4 & 21.4 \\ 
 &  &  &  & 21.8 \\ 
 &  &  &  & 22.2\\
\hline
\end{tabular}
\label{tab:input_par}
\end{table}

\subsubsection{Set of chemical abundances}

The metallicity in HII regions and in star-forming galaxies may be determined by direct methods (\cite{aller84}), measuring the line flux ratios, once the
temperature is determined, or through empirical but well tested metallicity sensitive calibrations (\cite{pilyugin2005}), when it is not possible to measure directly
the abundances (high metallicity or low excitation cases). In the stellar photoionization case, the ionization structure is known and so it is possible to obtain the
total abundances by measuring the ionic abundances using the ionization correction factors method (see (\cite{mathis91}) and references therein).
In the AGN, the situation is more difficult to handle since, when the ionizing spectrum is a no-thermal power-law, the ionization structure is very complex: 
the partial ionized region is so large so that different ionization states are mixed and taking into account the ionic abundances making up the total 
abundances gets quite complicated. In this case, direct methods are not possible and the empirical calibrations are obtained only by models. The choice of the 
sets of chemical abundances a critical step in the creation of the models because they are critically linked to another fundamental aspect of the simulations, i.e. 
the role of the dust and its composition. Once a reference sample of different abundances is assumed (usually, the solar set of values), a law describing the
change in the chemical composition and that will be used for the exploration of different metallicities, is required. The observations of the ionized gas chemical 
abundances in HII regions and starburst galaxies show that all the elements, except nitrogen and helium, in first approximation are found in equal 
proportions. As usual, in the HII regions the metallicity is expressed in terms of oxygen abundance while in the visible spectra it is possible 
to measure three different ionization states and oxygen is the most abundant element after hydrogen and helium. Then, we can assume that 
all the elements scale directly with the oxygen abundance, except for nitrogen and helium that must be treated separately. The nitrogen abundance, as well as 
helium, must be corrected for secondary production \cite{vce93}.

\begin{equation}
 \frac{N} {H} = \frac{O}{H} [10^{-1.6}+10^{2.37+log(O/H)}].
\end{equation}
\label{equ:Nsec}

\begin{equation}
\frac{He} {H} = 0.0737 + 0.0293 \frac{Z}{Z_{\odot}} 
\end{equation}

The composition and quantity of dust are also sources of uncertainties. The set of abundances needs to take into account both gas and dust composition in order 
to maintain the total assumed element abundances. As for the abundances of the elements, whose reference values are assumed to be similar to the solar 
values, the default values of abundance and composition for dust are taken from the Galactic interstellar matter (ISM). Chosen the features of the dust, the 
abundances of the gas component are calculated by subtracting the dust abundances from the total. With the exception of the noble gases, all abundances must 
be corrected for depletion into dust. This correction corresponds to a variation in the abundances of elements in gaseous form. The determination of these 
corrections is a main issue of modeling since it is only possible to make assumptions based on the observed properties of the Galactic ISM. It is reasonable to
suppose an increase of dust with metallicity because at a metallicity increase corresponds an increment of the refracting elements. Another important 
assumption regards the dust to gas ratio value (D/G thereafter). By changing the relative internal dust content as a function of the metallicity, the depletion factors 
must also be changed, and a method to estimate the depletion factors with the variation of the dust content is presented in (\cite{bin93}). In any case, some 
assumptions for the D/G and the variation of the dust with the metallicity are necessary. It seems that the depletion into grains could be constant in many 
directions of the Galaxy (see (\cite{vla02})), so we adopted the depletion factors reported in (\cite{groves2006}) as reference values and a linear law for the
scaling of the dust abundance with the metallicity. Using the conventional logarithmic notation, the relation between depletion factor 
and gas abundance is given by the formula:

\begin{equation}
depl(X)=log(X/H)_{gas}-log(X/H)_{tot}
\end{equation}

\begin{figure}
 \includegraphics[width=6cm]{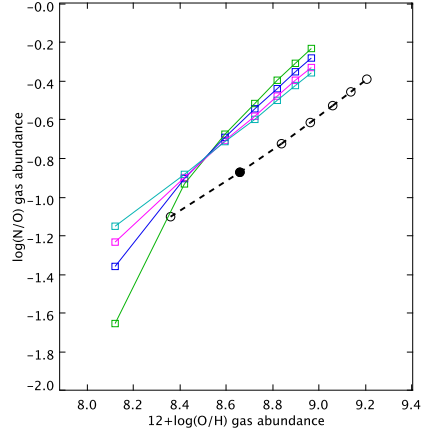}
 \caption{N/O vs O abundance, black dashed line total abundances, filled black circle the Sun abundances, colored lines: pale blue D/G=1.00, 
 cyan D/G=0.75, blue D/G=0.50, green D/G=0.25 }
 \label{fig:N_O_gas}
\end{figure}  

\begin{figure}
 \centering
 \includegraphics[width=6cm]{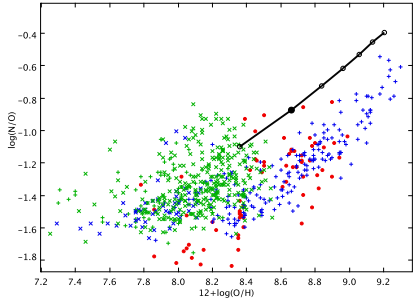}
 \caption{the log(N/O) vs 12+log(O/H): filled red circles UV galaxies, blue crosses HII regions in spiral galaxies, green crosses HII regions in blue compact 
 galaxies, blue *** HII regions in dwarf irregular galaxies and green *** emission line galaxies, black line the adopted N/O function, the black circles show the    
 adopted metallicity values $Z/Z\odot$, respectively 0.5, 1 (filled circle), 1.5, 2, 2.5, 3, 3.5.}
 \label{fig:N_O_collection}
\end{figure}

In this work, it is assumed $log(X/O)=$cost for all chemical elements, with the exception of N and He. For these elements we follow (\cite{gds04}) and (\cite
{groves06}) by employing a linear combination of the primary and secondary components of Nitrogen with the requirement of matching the adopted solar 
abundance patterns (\cite{asplund2005}) (Table \ref{tab:asp_ab}). In the models, $4$ D/G values were used in units of the ISM D/G = $0.25$, $0.50$, $0.75$, 
$1.00$. For each D/G value, we defined seven sets of chemical abundances, using the solar oxygen abundance as reference: $Z/ Z_{\odot}$=$0.5$; $1$; $1.5$; 
$2$; $2.5$; $3$; $3.5$.  We focused our attention on the high metallicity levels because nuclear gas and Seyfert galaxies very rarely present sub-solar 
metallicities (\cite{groves2006}). For each adopted D/G we changed the depletion factors set in order to retain the total abundances. The set of depletion factors 
is constant for each assumed D/G value with the exception of nitrogen because of the secondary production. The nitrogen depletion factor must change with the 
metallicity because, in agreement with our assumptions, if the dust increases with the metallicity there must be an excess of nitrogen in gas form due to the 
formula \ref{equ:Nsec}. The Nitrogen may be depleted in the carbonaceous grains: the involved functional groups are Amines (N-H, C-N), Amides (N-H), Nitriles 
(C,N triple bond stretch), Isocyanates (-N=C=O), Isothiocyanates (-N=C=S), Imines (R2C=N-R) and Nitro groups (-NO2 aliphatic, aromatic) \cite{kwok07}. 
Excluding the ammonia (NH$_{3}$), which cannot survive in an intense ionizing field such as in the AGN case, all the mentioned compounds need their partner 
elements linearly increasing with the metallicity. In Table \ref{tab:ndepl100} the depletion factors used for D/G$=1.00$ are reported. if this assumption is correct, 
a great increment of nitrogen gas abundance for $Z/ Z_{\odot} > 1.5$  should be observed, suggesting that it is not necessary to invoke an excess of nitrogen in 
order to match the observed $[NII]\lambda6584$ line intensity. In the Figure \ref{fig:N_O_gas} there are the relations between $log(N/O)_{gas}$ 
and $12+log(O/H)_{gas}$ for each couple of D/G and $Z/ Z_{\odot}$ values. For the other elements the trend is constant at a given D/G value. It is important to 
stress that the depletion of the elements must be taken into consideration even if the nebula is without dust, such as in radiation-pressure dominated models of 
(\cite{gds04}) according to which the dust is blown away by the wind. In our models, we assumed only two types of dust, graphite and silicates; their reference 
abundances are $[C/H]=1.22E-04$ when D/G$=1$ in the graphite and $[O/H]=1.94E-04$, $[Mg/H]=3.15E-05$, $[Si/H]=2.82E-05$, $[Fe/H]=2.70E-05$ for 
silicate respectively, while the distribution of their size is in the range $[0.005,0.25]$ $\mu$m. We analyzed the abundances found in many HII regions with direct 
methods in order to check the relation between oxygen abundance and the other elements. In (\cite{vz98}), (\cite{vzh06}), (\cite{izo99}), (\cite{izo06}) and 
(\cite{cont02}) a vast collection of chemical abundances were determined in $186$ $H II$ regions spanning a range of radius in $13$ spiral galaxies, $54$ 
supergiant H II regions in $50$ low-metallicity blue compact galaxies, $67$ $H II$ regions in $21$ dwarf irregular galaxies, $~310$ emission-line galaxies from 
the Data Release $3$ of the SDSS and a sample of $68$ UV-selected intermediate-redshift ($0\leq z \leq0.4$) galaxies. Our set of models does not match very
well with the trend seen in Figure \ref{fig:N_O_collection}, whereas there is a better fit with the samples used by (\cite{gds04}) (\cite{mou02},
\cite{ken03}) but the $log(N/O)$ trend is in any case overestimated when the abundances are greater than $Z_{\odot}$. In order to compare our results with 
those obtained by the above mentioned authors, we adopted the same solar abundances, formalism and constraints. It is apparent that the UV-galaxies have a 
large spread in the $log(N/O)$ values, so it could be possible that the HII regions and starburst galaxies have different sets of chemical abundances. This 
hypothesis could be tested by using a sample of gas abundances in nuclei of spiral galaxies to ascertain whether the adopted function is an upper limit 
of the real one. The final number of models models is $10390$, $9782$, $9022$ and $8207$ for D/G=$0.25$, $0.50$, $0.75$, $1.00$ respectively. The models 
will be rejected unless the resulting spectra are classified as Seyfert 2 spectra in the (\cite{veilleux}) diagnostic diagrams, following the Kewley's conditions 
(\cite{kewley2006}), and the synthetic fluxes are in the observed ranges, following Koski's diagrams (\cite{k78}).

\begin{table}
  \caption{Adopted abundances and depletion factors (the depletion factor is defined as:
 $depl=log(X/H)_{gas}-log(X/H)_{tot}$).}
  \begin{tabular}{@{}lcccccc}
  \hline
 el. & $log(X/H)_{\odot}$	& depl & $log(X/H)_{\odot gas}$ & ISM\\
\hline

   He    &-1.07  &0.00  &-1.07  &     \\
   Li    &-10.95  &0.00  &-10.95  &     \\
   Be    &-10.62  &0.00  &-10.62  &     \\
   B    &-9.30  &0.00  &-9.30  &     \\
   C    &-3.61  &-0.30  &-3.91  &-3.91\\
   N    &-4.22  &-0.30  &-4.52  &-4.52\\
   O    &-3.34  &-0.24  &-3.58  &-3.71\\
   F    &-7.44  &0.00  &-7.44  &     \\
   Ne    &-4.16  &0.00  &-4.16  &     \\
   Na    &-5.83  &-0.60  &-6.43  &-5.96\\
   Mg    &-4.47  &-1.15  &-5.62  &-4.50\\
   Al    &-5.63  &-1.44  &-7.07  &-5.65\\
   Si    &-4.49  &-0.89  &-5.38  &-4.55\\
   P    &-6.64  &0.00  &-6.64  &     \\
   S    &-4.86  &-0.34  &-5.20  &-5.13\\
   Cl    &-6.50  &-0.30  &-6.80  &-6.80\\
   Ar    &-5.82  &0.00  &-5.82  &     \\
   K    &-6.92  &0.00  &-6.92  &     \\
   Ca    &-5.68  &-2.52  &-8.20  &-5.68\\
   Sc    &-8.95  &0.00  &-8.95  &     \\
   Ti    &-7.10  &0.00  &-7.10  &     \\
   V    &-8.00  &0.00  &-8.00  &     \\
   Cr    &-6.36  &0.00  &-6.36  &     \\
   Mn    &-6.61  &0.00  &-6.61  &     \\
   Fe    &-4.55  &-1.37  &-5.92  &-4.57\\
   Co    &-7.08  &0.00  &-7.08  &     \\
   Ni    &-5.77  &-1.40  &-7.17  &-5.79\\
   Cu    &-7.79  &0.00  &-7.79  &     \\
   Zn    &-7.40  &0.00  &-7.40  &     \\
\hline
\end{tabular}
\medskip
\label{tab:asp_ab}
\end{table}

\begin{table}
  \caption{Nitrogen depletion factors for the ratio D/G=1.00. Relative abundances of metals are in logarithmic notation.}
  \begin{tabular}{@{}lcccccc}
  \hline
  $Z/Z\sun$ & $O/H$ & $N/H$ & $N/H_{gas}$ & depl\\
  \hline
0.5&-3.64&-4.75&-5.54&-0.79\\
1&-3.34&-4.22&-4.52&-0.3\\
1.5&-3.16&-3.89&-4.09&-0.19\\
2&-3.04&-3.66&-3.8&-0.14\\
2.5&-2.94&-3.48&-3.59&-0.11\\
3&-2.86&-3.32&-3.41&-0.09\\
3.5&-2.8&-3.19&-3.27&-0.08\\
\hline
\end{tabular}
\label{tab:ndepl100}
\end{table}

\subsubsection{Composite models}

The simple single cloud model described in the previous section is( only the first step of our modeling effort since in the observed Seyfert spectra two distinct c
components are usually observed, yielding high and low ionization lines. A double component model with two clouds, each accounting for either the high or 
the low ionization lines, can be created. We explored two kinds of composite models: Binette's models and 2 clouds models. The first one is based on the 
(\cite{bws96}) work, with the only improvements represented by the exploration of a larger region of the parameter space; the second model is based on a
combination of two different and independent single cloud models (\cite{ks97}), as described previously. We considered all the possible combinations between 
two clouds with fixed power-law indexes and metallicities. The spectral lines emitted by the first and second cloud are combined by a weighted mean, where the 
weight is given by the product of $H_{\beta}$ luminosities ratio and the solid angle ratio ($\omega_{1}/\omega_{2}$) subtended by 
the clouds as seen from the source of the ionizing radiation (this ratio, called geometrical factor, is indicated by $GF$). The following steps lead to the relation 
which has been applied to the combined fluxes.
The line luminosity is given by:

\begin{equation}
 L=4\pi r^{2} S \omega /4\pi
\end{equation}

\noindent where $r$ is the inner radius of the nebula, $\omega/4\pi$ is the covering factor, $S$ the emission line intensity ($erg$ $s^{-1}$ $cm^{-2}$). Then:

\begin {equation}
 L=4\pi d^{2} F=4\pi r^{2} S \omega /4\pi
\end{equation}

\noindent where $F$ is the observed flux ($erg$ $s^{-1}$ $cm^{-2}$), $d$ is the distance of the source.
Defining the density of ionizing photons ($s^{-1}$ $cm^{-2}$) as:

\begin{equation}
\phi (H^{0})=Q(H^{0})/4\pi r^{2}
\end{equation}

\noindent where $Q(H^{0})$ is the number of ionizing photons emitted from the source in a second and the ionization parameter as 

\begin {equation}
 U=\phi (H^{0})/n_{H} c
\end{equation}

\noindent If there are two clouds:

\begin{equation}
 U_{1}/U_{2}=r_{2}^{2} n_{2}/r_{1}^{2} n_{1} \Rightarrow r_{2}^{2}/r_{1}^{2}=U_{1} n_{1}/U_{2} n_{2}
\end{equation}

\noindent and, as a consequence:

\begin{equation}
 L_{2}/L_{1}=4\pi d^{2} F_{2}/4\pi d^{2} F_{1}=F_{2}/F_{1}
\end{equation}

\noindent Then:

\begin{equation}
F_{2}/F_{1}=U_{1} n_{1} S_{2} \omega_{2}/ U_{2} n_{2} S_{1} \omega_{1}
\end{equation}

\noindent in terms of $H_{\beta}$ this ratio can be written as:

\begin{equation}
F_{2} (H\beta) / F_{1} (H\beta)= GF \cdot L_{2}^{'}(H\beta)/L_{1}^{'}(H\beta)
\end{equation}

\noindent where $L_{2}^{'}$($H_{\beta}$)/$L_{1}^{'}$($H_{\beta}$) is the $H_{\beta}$  luminosity ratio if $\omega_{2}=\omega_{2}$.
These quantities are calculated by CLOUDY for each model. If we want to find the emission line intensity $I_{\lambda}$ 
relative to $H_{\beta}$ then:

\begin{equation}
I_{\lambda}=\frac{(F_{\lambda})_{2}+ (F_{\lambda})_{1}}{F_{tot} (H\beta)}
\end{equation}

\noindent where $F_{tot} (H\beta)$ is the total observed flux, and rearranging the terms 
we obtain:

\begin {equation}
I_{\lambda}=\frac{(I_{\lambda})_{2}+\frac{F(H\beta)_{1}}{F(H\beta)_{2}} (I_{\lambda})_{1}}{1+\frac{F(H\beta)_{1}}{F(H\beta)_{2}}}
\end{equation}

\noindent and finally:

\begin{equation}
 I_{\lambda}=\frac{ (I_{\lambda})_{2}+\frac{L_{2}^{'}(H\beta)}{L_{1}^{'}(H\beta)} \cdot GF \cdot (I_{\lambda})_{1}}{1+\frac{L_{2}^{'}(H\beta)}{L_{1}^{'}(H\beta)}\cdot GF}
\end{equation}
\label{equ:ave_ami}

\noindent If $GF$ goes to zero only the second cloud is visible; on the contrary, if $GF$ goes to infinity, only the first cloud is visible. Five values for 
$GF$ have been used ($0.25$, $0.5$, $1$, $2$ and $4$), so that each couple of models provides five new synthetical spectra.The total number of 
models obtained with the constrains on D/G, $\alpha$, $Z/Z_{\odot}$ fixed for each couple and after imposing the Kewley and Koski conditions, is 
about $9\cdot 10^{6}$. 

\begin {figure}
 \includegraphics[width=6cm]{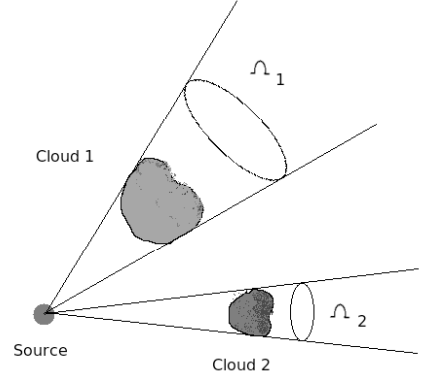}
 \caption{2 clouds model, $\omega_{1}$ and $\omega_{2}$ are the solid angles subtended by the clouds.}
 \label{fig:2clouds} 
\end {figure}

\subsection{Comparison of models with observations}

We have compared all the measured lines coming from the observations with the synthetic data provided by the models. Our goal is to reproduce as 
efficiently as possible the observed spectra with the aim of determining the most realistic set of input parameters. In (\cite{omm99}), the authors proposed 
a new method for deriving abundances in the AGN narrow line region (NLR), consisting in a selection of  "good'" models from a great sample (they used
$27000$ models) which is capable of fairly reproducing the observed line spectra, and then slightly modifying the chemical abundances in order to determine 
the best estimates of the metallicity of the AGN. The very large number of synthetical spectra we have produced has been employed to determine the set of 
models yielding the most similar synthetic spectral features to each observed spectrum with a $\chi^{2}$ test.
In this case, the $\chi^{2}$ is given by: 

\begin{equation}
 \chi^{2}=\sum_{i} (F_{\lambda_{obs}} - F_{\lambda_{C}})^{2}_{i}/(\sigma_{i})^{2}
\end{equation}

\noindent where $F_{\lambda_{obs}} - F_{\lambda_{C}}$ is the difference between the observed flux and CLOUDY flux, and $\sigma_{i}$ the error associated to 
the i-th line. The most probable set of input parameters can be therefore determined for each observed spectrum. The comparison has been carried out taking 
into account the synthetical lines which are foreseen to be visible in the observed spectrum, since these lines would be detectable ($S/N>3 $ ratio) in the observed 
spectrum. Lines below the $S/N$ value of the observed spectrum or not visible are excluded from the analysis. Two further parameters considered in the models 
are the reddening (Av) and the scale factor $F(H_{\beta})$. The $\chi^{2}$ test is performed using the measured fluxes not corrected by reddening while the 
flux models are reddened in order to achieve the observed  $F(H_{\alpha})/F_{(H\beta)}$ ratio. Since we have not corrected for the extinction using a fixed value, 
even if close to $3.1$, many models show a significantly different value. Av is calculated applying the relation in (\cite{ccm89}): 

\begin{equation}
 Av=7.22\cdotp \left[ log\left( \frac {F(H\alpha)} {F(H\beta)}\right) _{o} - log\left( \frac{F(H\alpha)}{F(H\beta)}\right) _{C}\right]  
\end{equation}

\noindent Nine different values of $F(H_{\alpha})/F(H_{\beta})$, obtained combining $F(H_{\alpha}) \pm \Delta F(H_{\alpha})$ with $F(H_{\beta}) \pm \Delta F(H_{\beta})$ 
(writing in compact form the fluxes $F_{\alpha}/F_{\beta}$, $(F_{\alpha}+ \Delta F_{\alpha})/F_{\beta})$, $F_{\alpha}/(F_{\beta}-\Delta F_{\beta})$ and so on), have been 
adopted. Then the mean ($\langle F_{\alpha}/F_{\beta} \rangle$) and the root mean square ($\sigma$) of the ratios are calculated, taking into account all the values within 
the range $-\sigma<\langle F_{\alpha}/F_{\beta}\rangle < \sigma$, giving as a result a positive Av value. For each model up to $9$ different synthetical spectra have been calculated.
The scale factor allows us to compare directly the reddened model with the observed spectrum. Relative fluxes would have increased relative errors because of the 
propagation. The accepted model have to satisfy a predetermined significance level, suggested by the analysis of the reduced $\chi^{2}$ distribution and given by:

\begin{equation}
 \chi^{2}/dof=\chi^{2}/(nr-np)
\end{equation}

\noindent where "dof" stands for degree of freedom, "nr" is the number of measured lines and "np" the number of parameters used in the models. The minimal number of dof necessary 
in order to decide the acceptability level is given by the number of measured lines minus the number of parameters used in the models. In order to consider the model a good 
approximation of the observed spectrum, the $\chi^{2}/dof$ distribution is required to be similar to a gaussian distribution peaked on $1$.  If the peak is located at values $<<1$
there is an overestimation in the fitting, presumably caused by the fact that very large error bars fit very well not significant set of parameters of the model. On the other hand, 
if the peak is placed at values $>>1$, the models cannot reproduce the observed spectra or the errors are too small. 
The significance level has been fixed, after several tests, between $5\%$ and $10\%$ and the assumed flux errors at $2\sigma$ level. The best fit models are associated to the 
minimum $\chi^{2}$ value, indicated with ($\chi^{2}_{min}$), with the assumption that if a model is accepted, all the models falling within the assumed confidence region are 
considered as acceptable as well. This condition can be expressed as:

\begin{equation}
 \chi^{2}-\chi^{2}_{min}<A
\end{equation}

where A is the value selected for a given probability \cite{mol02} ($25\%$ level of confidence is adopted in this work). For each fitted spectrum there is a group of accepted
models; the best values of the parameters are evaluated by averaging the single values over all the accepted models. In this analysis, $7$ free parameters for the single 
cloud models (SC) have been considered. As a consequence, a minimum number of measured lines equal to $7$ and $11$ 
to fit such models is required, while at least $12$ measured lines and $7$ parameters for Binette's models. About $50\%$ of the spectra shows less than $12$ measured lines, 
meaning that the 2C models can be compared to only the 50\% of the original sample of galaxies but with the great advantage of considering only the higher S/N spectra.
A spectrum with a low number of measured lines has usually low S/N, i.e. then big errors on the measured spectral lines, so, even though it is easier to fit the models with 
different parameters, these fits have little or no scientific validity. 

\subsubsection{The test}

In order to compare the different kinds of models, the spectra are splitted into two groups, namely spectra with $nr\geq12$ and spectra with $nr<12$. Since the D/G ratio is not 
considered a parameter of the model, the models with different D/G values will be treated separately. Such classification leads to four families of models, one for each 
D/G value. The number of input parameters for each kind of model are summarized in Table \ref{tab:para}.

\begin {table*}
\begin{center}
\caption{The input parameters of the simulations for single cloud, double clouds and Binette's models.}
\begin{tabular}{|l|c|c|c|}
\hline
Parameter & SC & 2C & Binette \\ 
\hline
power-law index, $\alpha$ & yes &yes &yes \\ 
metallicity, $Z/Z_{\odot}$ &yes &yes &yes \\ 
density, Ne &yes &yes &fixed \\ 
column density, Nc(H+)  &yes &yes &yes \\ 
ionization parameter, U &yes &yes &yes \\ 
density 2nd cloud, Ne$_{2}$ &no &yes &fixed \\ 
column density 2nd cloud, Nc(H+)$_{2}$ &no &yes &calculated \\ 
ionization parameter 2nd cloud, U$_{2}$ &no &yes &calculated \\ 
geometrical factor, Gf &no &yes &yes \\ 
Reddening, Av &yes &yes &yes \\ 
Scale factor, F(H$_{\beta}$) &yes &yes &yes \\ 
\hline
number of parameters &7 &11 &7\\
\hline
\end{tabular}
\label{tab:para}
\end{center}
\end{table*}

\subsubsection{Seyfert 2}

On a total of $2153$, $1141$ spectra have $nr<12$ and the remaining $nr\geq12$. For this reason, the comparison between SC and 2C models will be done 
only over $1012$ spectra. In Table \ref{tab:mod_res_sey2} the number of fitted spectra for each D/G value is reported; in more details, the first column contains the D/G value, 
second to fourth columns contain the fitted spectra with SC models and the last column the fitted spectra with 2C models. The total percentage of spectra which passed the test are
$\sim 65\%$ and $\sim 50\%$ for the SC and 2C models respectively. The percentage of spectra successfully fitted with $nr<12$ is $80\%$, while $\sim 45\%$ is the fraction 
of spectra fitted with $nr\geq12$. Both two 2C models accounts for only $50\%$ of the spectra fitted, since this work was carried out with the goal of achieving the best fitting 
and not the maximum number of spectra fitted. 

\begin{table*}
\begin{center}
\caption{Number of fitted spectra.}
 \begin{tabular}{|l|c|c|c|c|}
\hline
D/G & SC & SC $nr<12$  & SC $nr\geq12$  & 2C  \\
\hline
0.25&1249/2153&853/1141&396/1012&404/1012\\
0.50&1375/2153&912/1141&463/1012&486/1012\\
0.75&1445/2153&930/1141&515/1012&554/1012\\
1.00&1426/2153&939/1141&487/1012&561/1012\\
\hline
\end{tabular}
\label{tab:mod_res_sey2}
\end{center}
\end{table*}

\begin{figure}
 \includegraphics[width=6cm]{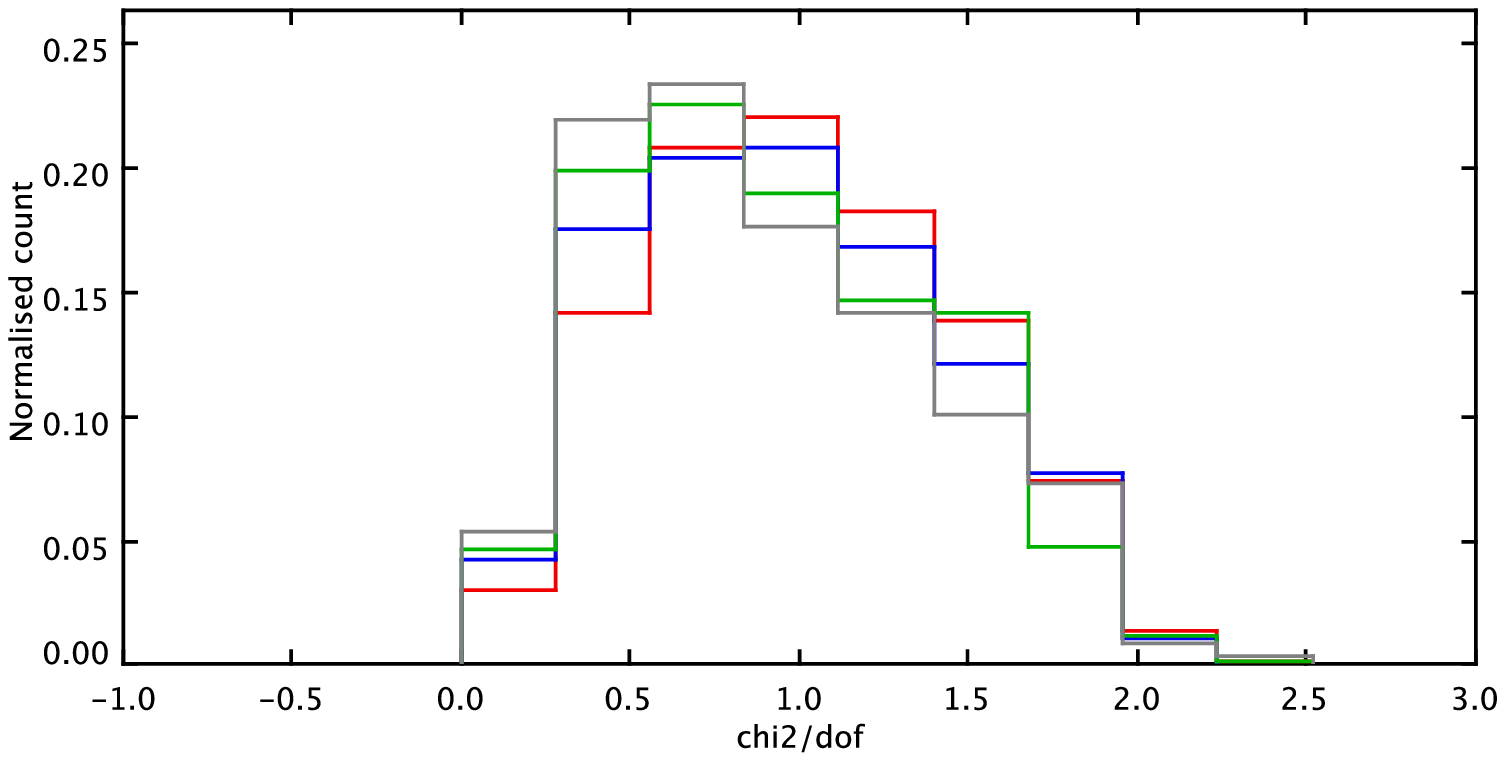}
 \includegraphics[width=6cm]{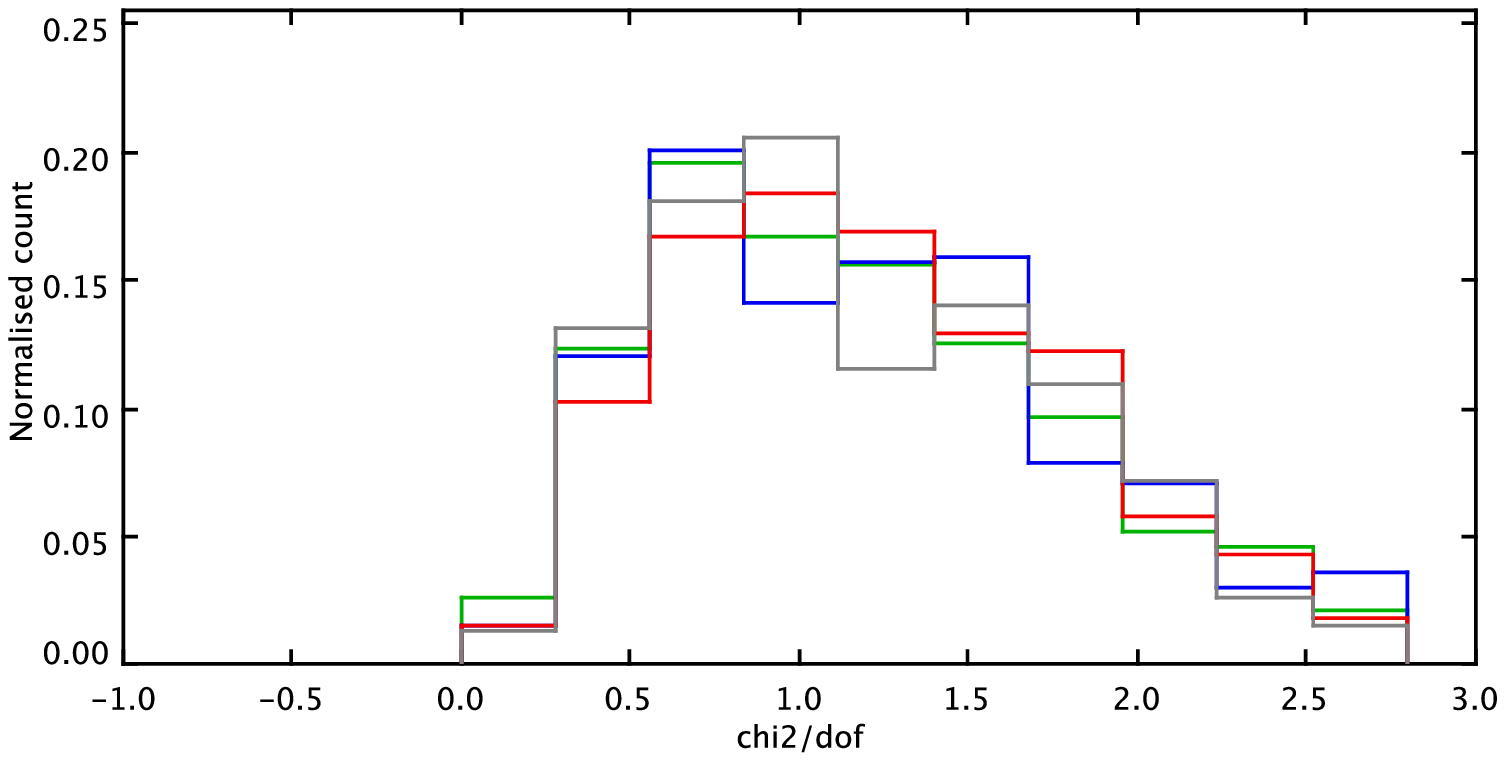}
 \caption{$\chi^{2}/dof$ distribution for SC models, up, and 2C models, down, and the different D/G values, red $0.25$, blue $0.50$, green $0.75$ and grey $1.00$.}
 \label{fig:chi2_dof_sey2}
\end{figure}

The $\chi^{2}/dof$ distributions are shown in figure \ref{fig:chi2_dof_sey2}. The distributions of the SC models are slightly skewed toward lower values for the D/G=$1$ and D/G=$0.75$ 
models, while the 2C models show tails for larger values even if their overall shapes are more regular. The significant difference between the models is evident 
comparing the distribution of absolute $\chi^{2}$, since this statistics represents the goodness of the overall fitting procedure (the input model parameters, the choice 
of the significance level and errors) and it is possible to discriminate between two models which are able to fit the same spectrum by looking at the $\chi^{2}$ value.

\begin{figure}
 \includegraphics[width=6cm]{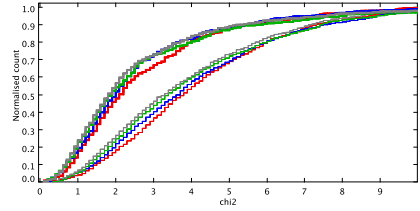}
 \caption{$\chi^{2}$ cumulative distribution for SC models, thin lines, and 2C models, thick lines.}
 \label{fig:chi2_cum_sey2}
\end{figure}

\begin{figure}
 \includegraphics[width=6cm]{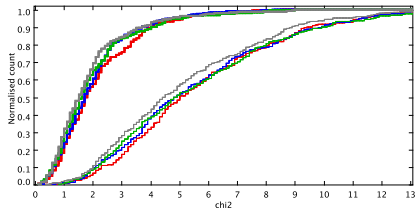}
 \caption{$\chi^{2}$ cumulative distribution for shared fitted spectra with the usual meaning of the symbols.}
 \label{fig:chi2_cum_sey2_shared}
\end{figure}

The cumulative distributions (Figure \ref{fig:chi2_cum_sey2}) of $\chi^{2}/dof$ show clearly the great difference in the obtained fits between the two kinds of models used 
in this work. Within each type of models, the D/G value yielding the minimum $\chi^{2}$ values is $1$ even if the difference is not significative, while the larger overall number of fitted 
spectra is obtained with D/G$=0.75 \div 1$. More interesting is the comparison of $\chi^{2}$  between the matched fitted spectra with single and 2C models. For each D/G value 
(starting from $0.25$) there are respectively $289$, $356$, $410$ and $396$ shared spectra, whose $\chi^{2}$ distributions are dramatically different. The differences can be 
appreciated looking at the cumulative distributions (Figure \ref{fig:chi2_cum_sey2_shared}): spectra having  good S/N are best fitted using 2C models. Many measured lines and 
much greater number of spectra could be fitted increasing the parameter resolution, in particular for the power-law index and the ionization parameter, at the cost of a much 
heavier computational load.

\noindent Table \ref{tab:line_fit} contains the percentage of failures for each measured line of the observed spectra. The data show again that the 2C models fit more efficiently all the lines.  
Anyway, these results should be cautiously handled since the percentage of successful fits are calculated averaging over few spectra, so that a single measure could modify drastically the 
result (in particular for the $[ArIV]$ and $[FeVII]$ lines). The lines mostly affected by this kind of issue are $[NI]5200$, $[FeVII]5721,6087$, $HeI 5876$ and $[ArIII]7135$. 

\begin{table*}
\begin{center}
\caption{Percentage of line failures for single cloud and double cloud models.}
\small
\begin{tabular}{|l|r|r|r|r|r|r|r|r|}
\hline
&  SC& & & & 2C &&&\\
\hline
 \multicolumn{1}{|c|}{line} &
  \multicolumn{1}{|c|}{0.25} &
  \multicolumn{1}{c|}{0.50} &
  \multicolumn{1}{c|}{0.75} &
  \multicolumn{1}{c|}{1.00} &
  \multicolumn{1}{c|}{0.25} &
  \multicolumn{1}{c|}{0.50} &
  \multicolumn{1}{c|}{0.75} &
  \multicolumn{1}{c|}{1.00} \\
\hline
  $[OII]$3727 & 6 & 6 & 7 & 5 & 3 & 2 & 2 & 1\\
  $[NeIII]$3869 & 2 & 2 & 1 & 1 & 2 & 0 & 0 & 3\\
  $[NeIII]$3968 & 1 & 1 & 2 & 3 & 5 & 2 & 1 & 1\\
  $[SII]$4074 & 8 & 18 & 18 & 25 & 0 & 0 & 0 & 0\\
  $H_{\gamma}$4340 & 0 & 1 & 1 & 1 & 0 & 0 & 0 & 0\\
  $[OIII]$4363 & 10 & 0 & 7 & 8 & 3 & 6 & 6 & 8\\
  HeII4686 & 2 & 1 & 1 & 3 & 0 & 1 & 0 & 0\\
  $[ArIV]$4711 & 0 & 0 & 0 & 0 & 0 & 0 & 0 & 0\\
  $[ArIV]$4740 & 33 & 25 & 25 & 20 & 0 & 0 & 0 & 0\\
  $H_{\beta}$4861 & 0 & 0 & 0 & 0 & 0 & 0 & 0 & 0\\
  $[OIII]$4959 & 0 & 0 & 0 & 0 & 0 & 0 & 0 & 0\\
  $[OIII]$5007 & 2 & 3 & 3 & 3 & 0 & 0 & 0 & 0\\
  $[NI]$5200 & 67 & 62 & 59 & 57 & 34 & 29 & 32 & 34\\
  $[FeVII]$5721 & 50 & 60 & 67 & 67 & 14 & 10 & 30 & 33\\
  HeI5876 & 19 & 17 & 16 & 15 & 8 & 10 & 9 & 11\\
  $[FeVII]$6087 & 62 & 61 & 66 & 63 & 9 & 15 & 15 & 65\\
  $[OI]$6300 & 7 & 6 & 7 & 6 & 2 & 3 & 5 & 5\\
  $[OI]$6363 & 2 & 1 & 1 & 1 & 0 & 0 & 1 & 0\\
  $[NII]$6548 & 0 & 0 & 0 & 0 & 0 & 0 & 0 & 0\\
  $H_{\alpha}$6563 & 0 & 0 & 0 & 0 & 0 & 0 & 0 & 0\\
  $[NII]$6584 & 2 & 3 & 4 & 5 & 0 & 0 & 0 & 0\\
  $[SII]$6716 & 2 & 3 & 3 & 4 & 0 & 0 & 1 & 1\\
  $[SII]$6731 & 4 & 4 & 4 & 4 & 8 & 6 & 4 & 1\\
  $[ArIII]$7135 & 7 & 11 & 16 & 22 & 3 & 8 & 26 & 27\\
  $[OII]$7325 & 0 & 0 & 0 & 0 & 0 & 0 & 0 & 10\\
\hline
\end{tabular} 
\label{tab:line_fit}
\end{center}
\end{table*}

\subsubsection{Metallicities of Seyfert 2 galaxies}

In Table \ref{tab:res_para_sey2} are summarized all the mean values of the parameters with their root mean square, $\sigma$, respectively for the single cloud and 2C models.
The power-law index $\alpha$ decreases to $-1.8$ with the increasing of the D/G ratio, probably because of the decreasing gas metallicity content (a large fraction of metals 
is depleted into grains). In fact, a low metallicity of the gas implies a much slower cooling and a weaker ionizing spectrum is necessary to obtain the same degree of ionization. 
The total metallicity is well defined in all models with negligible scatter; $1.5 \div 2.5$ $Z/Z_{\odot}$ is the range of the fitted values with the SC models while for the 2C model 
the estimate obtained is $1.0 \div 2.0$ $Z/Z_{\odot}$. The high metallicity of the gas indicates that the AGN fuel is evolved material, which, in turn, supports the idea that 
the gas has a local origin in the vast majority of the observed Seyfert 2 galaxies. The densities are in agreement with the measured values: SC and SC models produce the same 
values of densities, both very similar to the measured sulfur density.

\begin{table*}
\begin{center}
\caption{Parameters output}
\begin{tabular}{|l|r|r|r|r|r|r|r|r|}
\hline
One cloud models & $0.25$ && $0.50$ && $0.75$ && $1.00$&\\
\hline
  \multicolumn{1}{|c|}{Parameter} &
  \multicolumn{1}{c|}{mean} &
  \multicolumn{1}{c|}{$\sigma$} &
  \multicolumn{1}{c|}{mean} &
  \multicolumn{1}{c|}{$\sigma$} &
  \multicolumn{1}{c|}{mean} &
  \multicolumn{1}{c|}{$\sigma$} &
  \multicolumn{1}{c|}{mean} &
  \multicolumn{1}{c|}{$\sigma$} \\
\hline
    $\alpha$ & -1.5 & 0.3 & -1.6 & 0.3 & -1.7 & 0.3 & -1.8 & 0.3\\
    $Z/Z_{\odot}$ & 2.0 & 0.6 & 1.9 & 0.6 & 1.9 & 0.5 & 1.8 & 0.5\\
    $log(Ne)$ & 2.3 & 0.7 & 2.3 & 0.7 & 2.2 & 0.6 & 2.1 & 0.7\\
    $log(Nc(H+))$ & 20.5 & 0.7 & 20.3 & 0.6 & 20.0 & 0.5 & 19.9 & 0.4\\
    U & -3.1 & 0.2 & -3.1 & 0.2 & -3.0 & 0.2 & -3.0 & 0.3\\
    Av & 1.4 & 0.6 & 1.5 & 0.6 & 1.5 & 0.6 & 1.6 & 0.6\\
\hline
2C models\\
\hline
    $\alpha$ & -1.3 & 0.3 & -1.4 & 0.3 & -1.6 & 0.3 & -1.8 & 0.3\\
    $Z/Z_{\odot}$ & 1.6 & 0.5 & 1.7 & 0.4 & 1.7 & 0.4 & 1.7 & 0.4\\
    $log(Ne)$ & 2.2 & 0.8 & 2.2 & 0.8 & 2.1 & 0.7 & 2.2 & 0.8\\
    $log(Nc(H+))$ & 20.2 & 0.4 & 20.2 & 0.4 & 20.2 & 0.4 & 20.2 & 0.4\\
    U & -2.7 & 0.4 & -2.6 & 0.4 & -2.6 & 0.4 & -2.5 & 0.4\\
    $log(Ne)_{2}$ & 3.0 & 0.9 & 2.8 & 0.9 & 2.7 & 0.9 & 2.5 & 0.8\\
    $log(Nc(H+))_{2}$ & 19.7 & 0.4 & 19.7 & 0.3 & 19.7 & 0.3 & 19.7 & 0.3\\
    U$_{2}$ & -3.3 & 0.2 & -3.3 & 0.2 & -3.3 & 0.2 & -3.3 & 0.3\\
    Av & 1.2 & 0.4 & 1.2 & 0.4 & 1.3 & 0.4 & 1.4 & 0.4\\
\hline
\end{tabular}
\label{tab:res_para_sey2}
\end{center}
\end{table*}

\begin{figure}
 \includegraphics[width=6cm]{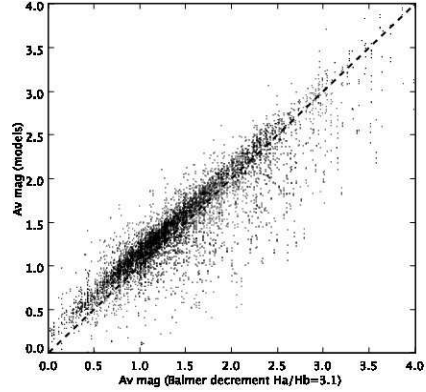}
 \caption{Comparison between the observed values of the absorption Av and the values produced by CLOUDY.}
 \label{fig:Av_bal_mod}
\end{figure}

The column densities are very similar in the SC models and in the first cloud of the 2C models, while the second cloud of the latter model has a lower column density and a lower 
ionization parameter too, which allows us to treat such cloud as a ionization bounded case. The ionization parameter in the SC models is almost constant for different D/G values: 
$log(U)=-3$. In the 2C the high ionization cloud shows a little variation in the ionization parameter with respect to D/G values, the mean value being $log(U)=-2.6$, while a very similar value 
is found for the low ionization cloud ($log(U)=-3.3$). In general, the amount of ionization in the SC models is intermediate between the values found in the first and second clouds 
of the 2C models. Another result, which can be explained with an increasing dust content, is the increase of the absorption Av with D/G. There is an average difference of $0.2$ mag 
between the SC and 2C models. In Figure \ref{fig:Av_bal_mod} the values of the Av found by models for all the fitted spectra and obtained with all the models are shown against the 
Av values estimated by Balmer decrement. The Av models appear to be slightly larger then the estimated values for both SC and 2C models. This effect can be explained by the fact 
that the $H_{\alpha}/H_{\beta}$ ratio obtained by CLOUDY, in the vast majority of cases, is significantly lower than the canonical value of $3.1$. 

\noindent The temperatures obtained by the models are shown in Table \ref{tab:temperatures}. The $[OII]$ and $[SII]$ temperatures values are consistent with those measured 
by RO2 and RS2t ratios, while the $[OIII]$ temperature is lower than the measured one but nonetheless consistent with the photoionization case. 

\begin{table*}
\begin{center}
\caption{Model temperatures for the observed and simulated spectral lines in single cloud and double clouds models.}
\begin{tabular}{|l|r|r|r|r|r|r|r|r|}
\hline
 & $0.25$ && $0.50$ && $0.75$ && $1.00$&\\
\hline
  \multicolumn{1}{|c|}{SC} &
  \multicolumn{1}{c|}{Te} &
  \multicolumn{1}{c|}{$\sigma$} &
  \multicolumn{1}{c|}{Te} &
  \multicolumn{1}{c|}{$\sigma$} &
  \multicolumn{1}{c|}{Te} &
  \multicolumn{1}{c|}{$\sigma$} &
  \multicolumn{1}{c|}{Te} &
  \multicolumn{1}{c|}{$\sigma$} \\
\hline
     ${[OIII]}$ & 8274 & 1421 & 8522 & 1437 & 8879 & 1387 & 9374 & 1379\\
     ${[OII]}$& 7813 & 1145 & 8142 & 1184 & 8568 & 1130 & 9136 & 1192\\
     ${[SII]}$& 7026 & 1038 & 7282 & 1163 & 7608 & 1220 & 8021 & 1527\\
\hline
2C\\
\hline
     ${[OIII]}$& 10923 & 3257 & 11172 & 3189 & 11225 & 2937 & 11535 & 2853\\
     ${[OII]}$& 10020 & 3353 & 10390 & 3169 & 10361 & 2806 & 10736 & 2658\\
     ${[SII]}$& 9219 & 3715 & 9560 & 3567 & 9334 & 3294 & 9759 & 3170\\
\hline
    ${[OIII]}_2$ & 9282 & 1118 & 9298 & 1085 & 9516 & 1103 & 9679 & 1140\\
    ${[OII]}_2$ & 8788 & 1078 & 8906 & 1016 & 9222 & 1020 & 9464 & 1006\\
    ${[SII]}_2$ & 7995 & 1418 & 8049 & 1353 & 8307 & 1450 & 8450 & 1480\\
\hline\end{tabular}
\label{tab:temperatures}
\end{center}
\end{table*}

\subsection{Metallicity for star-forming galaxies}

The determination of chemical abundances in the HII regions of SFs galaxies is complicated by the difficulty of reliable measures of the temperature of the HII clouds. 
The auroral lines, such as [OIII]4363, [SIII]6312 and [NII]5755, used to evaluate the temperature are usually weak and in low-excitation metal-rich HII regions they are 
often too weak to be detected. For this reason, the direct method is almost useless in the vast majority of the cases. The paper by (\cite{pagel1979}) suggested that some 
empirical calibrations, based on collection of observations, between the nebular line fluxes could be used to derive directly the oxygen abundance or the temperature of 
the HII regions. In more details, the parameter defined as $R_{23}=\frac{([OII]3727 + [OIII]4959,5007)}{H\beta}$ was used to directly measure the Oxygen abundances. 
This ratio was recalibrated using photoionization models and confirmed through other observations. Nonetheless, the comparison between the results obtained from 
different samples of galaxies remains very difficult because of the high heterogeneity of the model parameters and data. In all cases, the one-dimensional calibrations 
are systematically wrong as showed in (\cite{pilyugin2005}). The only $R_{23}$ index does not remove all the degenerations, so a more general two-dimensional 
parametric calibration, called the P-method, was proposed (\cite{pilyugin2005}), introducing a new parameter, called the excitation parameter and defined as:

\begin{equation}
P_{23}=\frac{([OIII]4959,5007)}{([OII]3727+[OIII]4595,5007)}
\end{equation}

From a collection of high precision measurements of 700 HII regions, in (\cite{pilyugin2005}) the authors revised the P-index calibration proposing two different functional forms for the 
upper branch and lower branch in the plane $R_{23}$ vs $\log{O/H}$ respectively. For the upper branch:

\begin{eqnarray*}
\lefteqn{12+\log{O/H} =} \\
& & \frac{(R_{23} + 726.1+842.2\cdot P + 337.5\cdot P^2)}{(85.96 + 82.76\cdot P +43.98\cdot P^2 + 1.793\cdot R_{23})}
\end{eqnarray*}

\begin{figure}
 \centering
 \includegraphics[width=6cm]{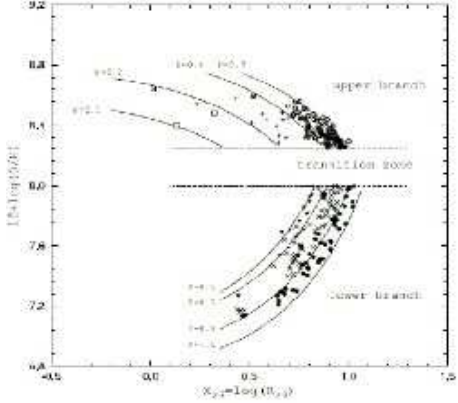}
 \caption{Family of curves in the $(R_{23}, \log{O/H})$ plane labeled by different values of the excitation parameter
                 $P$, superimposed with data from different observations. The high-metallicity HII regions with $0.0<P<0.3$ are 
                 shown as opens squares, those with $0.3<P<0.6$ as plus sign and those with $0.6<P<0.9$ as open circles.
                 The low-metallicity HII regions with $0.5<P<0.7$ are shown as filled triangles, those with $0.7<P<0.9$ as crosses and
                 those with $0.9<P<1.0$ as filled circles.}
 \label{fig:diagnostic}
\end{figure}

\noindent These relations can be used to determine the Oxygen abundances of the HII regions in SF galaxies and remove the degeneracies, since the P-index increases from left to 
right in the figure \ref{fig:diagnostic}.

\subsubsection{Our sample}

\begin{figure}
 \centering
 \includegraphics[width=6cm]{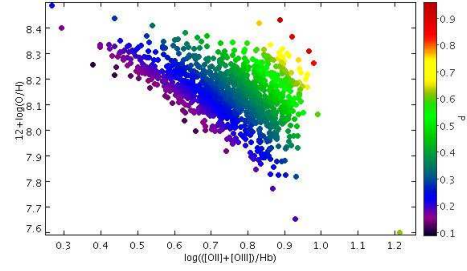}
 \caption{The star-forming galaxies of the sample used in this work in the $(R_{23}, \log{O/H})$. These galaxies have excitation parameter
                 $P$ in the interval $[0.1, 0.9]$, values that are consistent with their location in the upper branch sequence of the plot.}
 \label{R23_O_P}
\end{figure}

\noindent Applying this procedure, we have found that our sample of SFs galaxies has P values in the range [0.1-0.9], consistent with the upper branch sequence and that the sources 
are in the transition zone \ref{R23_O_P}. At the same time, the estimated Oxygen abundance of our sample is in the range [7.9-8.4]. Assuming the oxygen solar abundance 
of $\frac{N(O)}{N(H)} = 4.57 10^{-4}$ (\cite{asplund2005}), the gas in this sample of galaxies has a sub-solar abundance. If the metal abundances scale with the oxygen 
abundance, how it is usually assumed, the gas metallicity is between 0.2 and 0.5 $Z_{\sun}$ in figure \ref{R23_O_P}. It needs to be stressed that these determinations of the 
metal abundances take into account only the gas component and that, in order to achieve the real metallicity, dust composition has to be carefully modeled and considered. 
Assuming that the ISM has solar abundance, about $40\%$ of the oxygen resides in silicate grains, implying that $\frac{N(O)}{N(H)} =1.94 10^{-4}$ oxygen atoms per hydrogen 
atom are in dust form (\cite{groves2006}). Even if there is an increase in the oxygen abundance, the metallicity is  in any case sub-solar or solar (the estimated value represents 
an upper limit because the dust content, usually, increases with the metallicity).

\begin{figure}
 \centering
 \includegraphics[width=6cm]{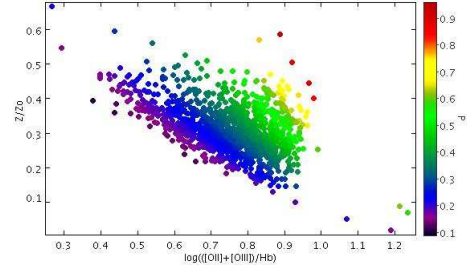}
 \caption{Same plot as before, with the assumption that total metallicity scales linearly with the Oxygen abundance. In this case, the SFs galaxies
                 of our sample have metallicity between $0.2\cdot Z_{\sum}$ and $0.5\cdot Z_{\sun}$}
 \label{R23_Z_P}
\end{figure} 

\section{Metallicity from stellar population synthesis}

The global relationship between the stellar populations and structural properties of hot galaxies has been studied in (\cite{bender1993})
using two different measures of the global stellar population, namely the $Mg_{2}$ index and the $(B-V)_{0}$ colour. The $Mg_{2}$ index,
introduced in (\cite{faber1977}), is defined as the the difference in magnitude between the instrumental flux in a window $F_{2} = [5156.0, 5197.25] \AA$
centered on the Magnesium spectral feature and the pseudo-continuum interpolated from two windows, $F_{1} = [4897.0, 4958.25] \AA$ and
$F_{3} = [5303.0, 5366.75] \AA$ at the blue and red sides respectively:

\begin{equation}
Mg_{2} = \frac{F_{2}}{F_{1} + 0.61(F_{3}-F_{1})}
\label{mg2}
\end{equation} 

Using population synthesis models of various authors, the relation between the $Mg_{2}$ index and the central velocity dispersion $\sigma_{0}$ can be translated
in an approximate relation between the same spectroscopic index and the age and metallicity of the stellar population:

\begin{equation}
\log{Mg_{2}} = 0.41\cdot \log{\frac{Z}{Z_{\sun}}}+0.41\cdot \log{t} -1.00
\label{rel_mg2}
\end{equation} 

where $Z$ is the metallicity, $t$ is the age measured in Gyr and the uncertainty on the coefficients is about $\pm 20\%$. This relation is relatively reliable for $-0.5 < \log{(Z/Z_{\sun})} <
0.3$ and $5< t <15$ (always expressed in Gyr). The age and metallicities of the stellar populations of the normal galaxies considered in our work have been determined 
by exploiting this relation. In turn, one of the two parameters has been derived from the fitted model of the continuum component of the spectrum obtained by STARLIGHT 
(whose output consist in a set of weights expressing the relative contribution of each single stellar population of given age and $Z$ to the observed spectrum); this estimate of
either $Z$ or $t$ has been used to calculate the other parameter as an unknown term in the \ref{rel_mg2} relation, using as approximation of the measured values of the $Mg_{2}$ values
the equivalent width of the MgI lines, retrieved from the SDSS database.  Both temperature and metallicity of the stellar population could have been, in principle, derived by the only
spectral models obtained by STARLIGHT, but this approach may prove to be essentially wrong due to the intrinsic limited coverage of the $(t, Z)$ plane of the library of template stellar
population used by STARLIGHT to fit the continuum component of the observed spectra, since every possible final estimates of both parameters is a weighted average of age
and metallicities of the stellar population contained in the template library.

\begin{figure}
 \centering
 \includegraphics[width=6cm]{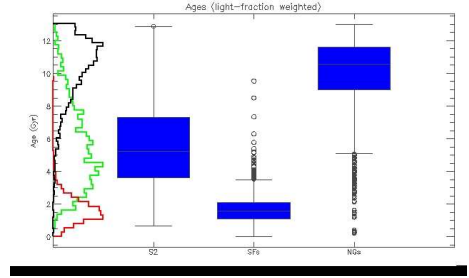}
 \caption{Distribution of mean weighted age $t$ for the three samples of galaxies considered in this work, obtained using STARLIGHT. The "box and whisker" plots show the position of the first four
                 quartiles of the age distribution and the outliers.}
 \label{starlight_t}
\end{figure} 

\begin{figure}
 \centering
 \includegraphics[width=6cm]{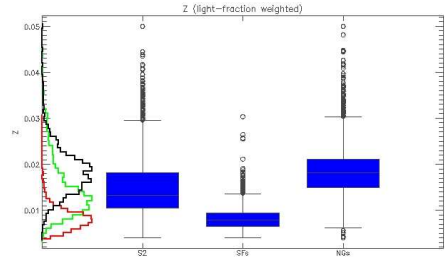}
 \caption{"Box and whisker" plots of the distributions of the mean weighted metallicity $Z$ for the three samples of galaxies considered in this work.}
 \label{starlight_Z}
\end{figure} 

\begin{figure}
 \centering
 \includegraphics[width=6cm]{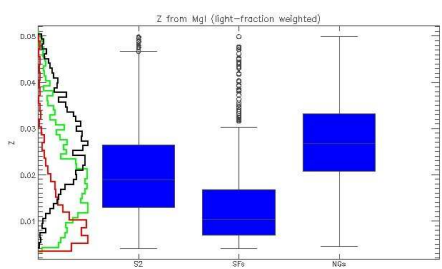}
 \caption{"Box and whisker" plots of the distributions of the mean weighted metallicities $Z$ for the three samples of galaxies considered in this work. These estimates
                 have been obtained using the relation \ref{rel_mg2}, the mean weighted age provided by STARLIGHT and the observed Mg2 indexes of the galaxies.}
 \label{mg2_Z}
\end{figure}

The age $t$ and metallicity $Z$ distributions for all galaxies belonging to the three different samples considered in this work and derived by STARLIGHT only are shown in figures 
\ref{starlight_t} and \ref{starlight_Z} as "whisker and box" plots with, on a side, the histograms obtained from the same distributions. As expected, the normal galaxies have significantly
older stellar populations than the other two samples of galaxies, among which SFs are still younger than Seyfert 2 galaxies. The metallicity distributions show that normal galaxies and 
Seyfert galaxies, while almost consistent from a statistical point of view, possess more evolved stellar populations than star-forming galaxies. On the other hand, the same distributions 
obtained using the measured values of the $Mg_{2}$ index and the relation \ref{rel_mg2} are shown in figures \ref{mh2_t} and \ref{mg2_Z} respectively. The comparison between these 
plots indicates that, though the relative relations between distributions of different samples of galaxies remain unchanged, stellar populations appear to be systematically younger and 
less metal-rich than in the case of the estimation by STARLIGHT only.

\begin{figure}
 \centering
 \includegraphics[width=6cm]{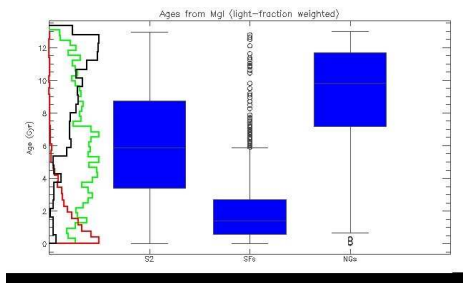}
 \caption{"Box and whisker" plots of the distributions of the mean weighted age $t$ for the three samples of galaxies considered in this work. These estimates 
                 have been obtained using the relation \ref{rel_mg2}, the mean metallicity provided by STARLIGHT and the observed Mg2 indexes of the galaxies.}
 \label{mg2_t}
\end{figure} 

\section{Conclusions}

In this work the metal content of the stellar and gaseous components of the central regions of galaxies has been characterized with multiple methods. Three different samples of
galaxies, namely normal galaxies, star-forming galaxies and galaxies hosting an AGN (Seyfert 2), retrieved from the SDSS DR7 spectroscopic database and classified according to an original 
method based on spectroscopic diagnostics (see \cite{vaona2009}), have been compared. For what the nuclear stellar populations are concerned, the presence of a strong blue continuum 
and the absence of any 4000\AA \ \ break in the spectra of the SF galaxies are an indication of the presence, in their circumnuclear regions, of an excess of young hot blue stars compared 
with NG and S2G. On the contrary, the spectra of the circumnuclear regions of S2G and NG indicate quite similar conditions: a dominating stellar component of type F and a 4000\AA \ \ break 
in NG deeper than in S2G of a factor $\sim$1.5. This indicates that the presence of an excess of A stars in the central regions of S2G compared to NG is quite likely, suggesting that 
there have been in the recent past of S2 AGN slightly more star formation events than in NG. The analysis of the gas of the central regions of the Seyfert 2 galaxies has been performed 
using a very robust modeling of the photoionization processes occurring in gas clouds around the AGN and exploring a very large region of the model parameters space in order
to reproduce realistically the observed spectra from simulation. Both single and two clouds models indicate for S2 a significantly larger metallicity than solar ($1.5 \div 2.5\ Z/Z_{\odot}$ with SC models 
and $1.0 \div 2.0\  Z/Z_{\odot}$ for the 2C models). These results support the idea that the AGN fuel is composed by evolved material and that the gas has a local origin in most of the observed Seyfert 2 galaxies.  The gas metal abundances for SF galaxies have been estimated using the method of the P-index (\cite{pilyugin2005}) and we have obtained results which clearly indicate sub-solar metal abundances (0.2 - 0.5 $Z_{\sun}$), confirming that young stars are fueled by relatively primordial and unprocessed material. The overall relations between the average age and metallicity of central stellar
populations of the three different samples of galaxies here considered have been determined using the $Mg_{2}$ absorption feature as proposed by(\cite{faber1977}) and exploiting the spectral continuum components
of observed spectra of the galaxies extracted using STARLIGHT (\cite{rafanelli2009}). This analysis qualitatively confirms that Seyfert 2 galaxies have young low-metallicity circumnuclear stellar populations 
feeding in a metal-rich environment where gas clouds are composed of rather high metallicity material; on the other hand, SF galaxies have very young and primordial central stellar populations. 
In conclusion, the comparative study of metallic content of stars and gas in this work indicate that Seyfert 2 nuclei occur preferentially in a stellar environment composed of young stars where recent 
circumnuclear star formation processes were active (as witnessed by the high metallicity of gas clouds), even if there is no compelling evidence that would connect present or past star formation 
activity to the nature of the central engine. The two phenomena, SF and AGN activity, may therefore be not causally connected to each other or, if so, through indirect or/and not clear physical mechanisms.

\bibliographystyle{aa}

\end{document}